\documentclass[Physsubmission, Phys]{SciPost}

\hypersetup{
    colorlinks,
    linkcolor={red!50!black},
    citecolor={blue!50!black},
    urlcolor={blue!80!black}
}

\usepackage[bitstream-charter]{mathdesign}
\DeclareSymbolFont{usualmathcal}{OMS}{cmsy}{m}{n}
\DeclareSymbolFontAlphabet{\mathcal}{usualmathcal}

\begin{document}

\begin{center}{\Large \textbf{
ISAI: Investigating Solar Axion by Iron-57
}}\end{center}

\begin{center}
Tomonori Ikeda\textsuperscript{1$\star$},
Toshihiro Fujii\textsuperscript{2},
Takeshi Go Tsuru\textsuperscript{1},
Yuki Amano\textsuperscript{1},
Kazuho Kayama\textsuperscript{1},
Masamune Matsuda\textsuperscript{1},
Hiromu Iwasaki\textsuperscript{1},
Mizuki Uenomachi\textsuperscript{1},
Kentaro Miuchi\textsuperscript{3},
Yoshiyuki Onuki\textsuperscript{4},
Yoshizumi Inoue\textsuperscript{4} and
Akimichi Taketa\textsuperscript{5}
\end{center}

\begin{center}
{\bf 1} Graduate School of Science, Kyoto University, Kitashirakawa Oiwake-cho, Sakyo-ku, Kyoto-shi, Kyoto, 606-8502, Japan
\\
{\bf 2} Graduate School of Science, Osaka Metropolitan University, Sugimoto-cho, Sumiyoshi-ku, Osaka-shi, Osaka, 558-8585, Japan
\\
{\bf 3} Department of Physics, Graduate School of Science, Kobe University, Rokkodai-cho, Nada-ku, Kobe-shi, Hyogo, 657-8501, Japan
\\
{\bf 4} ICEPP, The University of Tokyo, 7-3-1 Hongo, Bunkyo-ku, Tokyo 113-0033, Japan
\\
{\bf 5} ERI, The University of Tokyo, 1-1-1 Yayoi, Bunkyo-ku, Tokyo 113-0032, Japan
\\
* ikeda.tomonori.62h@st.kyoto-u.ac.jp
\end{center}

\begin{center}
\today
\end{center}


\definecolor{palegray}{gray}{0.95}
\begin{center}
\colorbox{palegray}{
  \begin{tabular}{rr}
  \begin{minipage}{0.1\textwidth}
    \includegraphics[width=30mm]{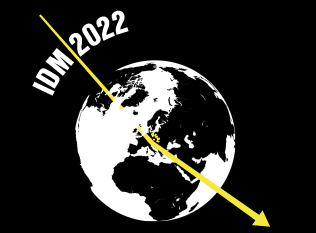}
  \end{minipage}
  &
  \begin{minipage}{0.85\textwidth}
    \begin{center}
    {\it 14th International Conference on Identification of Dark Matter}\\
    {\it Vienna, Austria, 18-22 July 2022} \\
    \doi{10.21468/SciPostPhysProc.?}\\
    \end{center}
  \end{minipage}
\end{tabular}
}
\end{center}

\newcommand{\ikeda}[1]{\textcolor{black}{#1}}

\section*{Abstract}
{\bf

The existence of the axion is a unique solution for the strong CP problem, and the axion is one of the most promising candidates of the dark matter. Investigating Solar Axion by Iron-57 (ISAI) is being prepared as a complemented table-top experiment to confirm the solar axion scenario. Probing an X-ray emission from the nuclear transitions associated with the axion-nucleon coupling is a leading approach. ISAI searches for the monochromatic 14.4\,keV X-ray from the first excited state of $^{57}$Fe using a state-of-the-art pixelized silicon detector, dubbed XRPIX, under an extremely low-background environment. We highlight scientific objectives, experimental design and the latest status of ISAI.
}

\section{Introduction}
\label{sec:intro}
Quantum chromodynamics requires extremely fine-tuning to explain the observed electric dipole moment of the neutron~\cite{Crewther:1979pi,Afach:2015sja}. This problem is generally known as the strong CP problem. Peccei and Quinn proposed a global \textit{U}(1) symmetry whose breaking gives a rise to a new pseudoscalar boson called axion~\cite{PhysRevLett.38.1440,PhysRevLett.40.223,Wilczek:1977pj}. Moreover, the axion could be a part of the dark matter~\cite{PhysRevLett.50.925}. Therefore, the axion search is one of the most important topics in the fields of astrophysics and particle physics.
The axion mass ($m_{a}$) and the coupling strength are proportional to the inverse of the breaking scale ($f_{a}$). The axion mass can be written with the following relation:
\begin{equation}\label{eq:mass}
    m_{a} = \frac{\sqrt{z}}{1+z}\frac{1.3\times 10^{7}}{f_{a}/{\rm{GeV}}} ~{\rm{eV}},
\end{equation}
where $z=m_{u}/m_{d}$ is the mass ratio of the up and down quark.
Since the higher breaking scale than the electroweak scale makes the weakly interacting, axions become invisible. 
One of the invisible axion models is the Kim-Shifman-Vainstein-Zakharov (KSVZ) axion, or called as the hadronic axion, which does not have tree-level couplings to leptons and ordinary quarks. 
The brightest source of such axion is the Sun, and the solar axion flux on Earth is calculated in Ref~\cite{Redondo_2013}. 
Some experiments have been performed as the axion helioscope and searched for the axion-photon and axion-electron coupling using the Primakoff effect or Compton interaction~\cite{CAST2017,PhysRevD.102.072004}.
The other novel method to independently verify the axion-nucleon coupling of the hadronic axion was proposed by Moriyama~\cite{PhysRevLett.75.3222}.
$^{57}$Fe nuclei are thermally excited in the Sun and can decay by emitting monochromatic axions corresponding to 14.4\,keV. Emitted monochromatic axions can resonantly excite $^{57}$Fe nuclei thanks to the doppler broadening of the axion energy due to the thermal motion of the $^{57}$Fe nuclei in the Sun. 
The expected excitation rate per unit mass of $^{57}$Fe is written as,
\begin{equation}\label{eq:rate}
    R= 3.0\times 10^{2}~\biggl( \frac{10^{6} {\rm GeV}}{f_{a}} \biggr)^{4}~C^{4} ~{\rm day^{-1}kg^{-1}},
\end{equation}
where $C$ depends on the nuclear structure.
The most stringent upper limit on the mass of the axion was reported in 2007 to be $m_{a} < 216$\,eV using the Si PIN photodiodes~\cite{NAMBA2007398}.
In this paper, we propose to search for the monochromatic 14.4\,keV axion with a pixelized silicon detector called XRPIX. Furthermore, we describe the detector design, its performance, and the predicted sensitivity to the mass of the axion.

\section{Detector design and instruments}
\subsection{Investigating Solar Axion by Iron-57 (ISAI)}
\begin{figure}[!h]
\centering
\includegraphics[width=\textwidth]{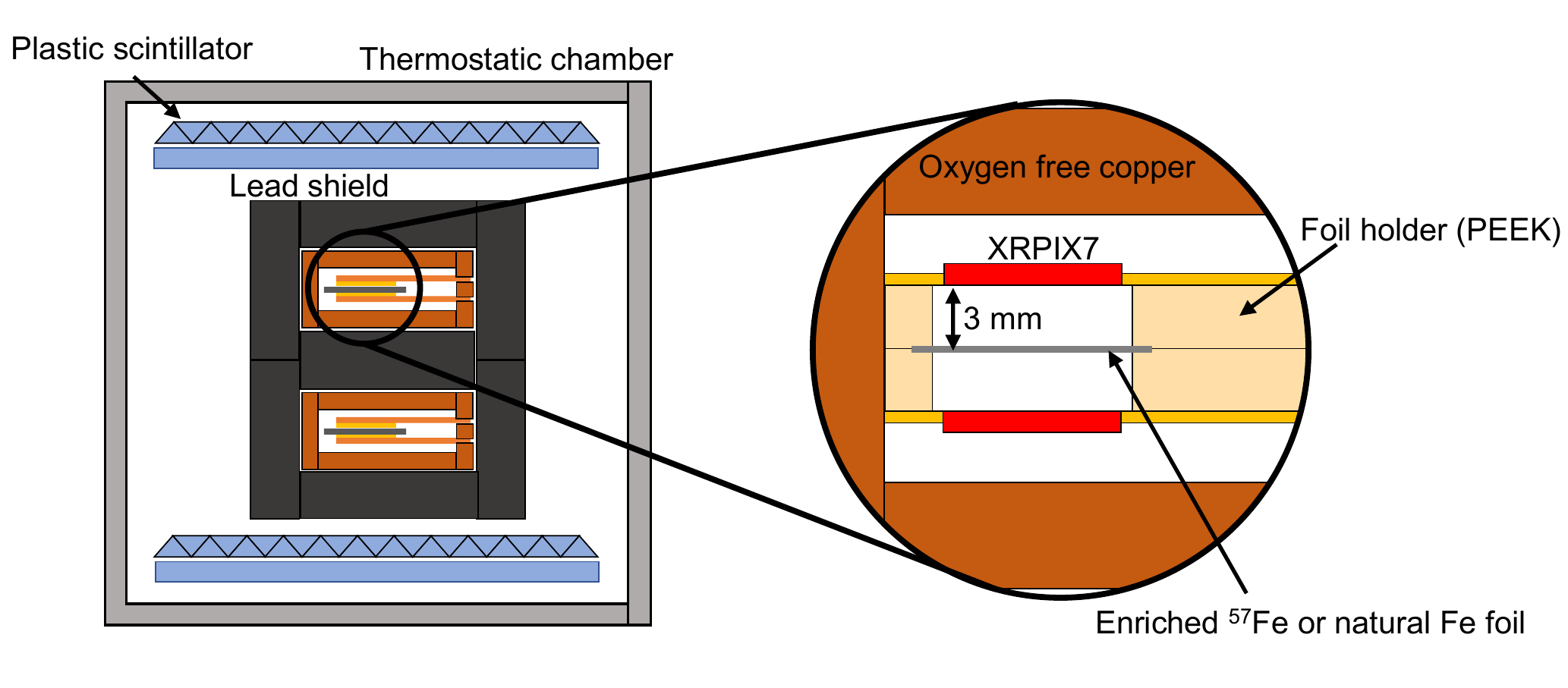}
\caption{Schematic drawing of the detector design of ISAI. The left picture shows the overall detector, which consists of two sets of main detector modules surrounded by plastic scintillators as the anti-counter and a thermostatic chamber. The right picture shows the details around the enriched $^{57}$Fe or natural Fe foil.}
\label{design}
\end{figure}

The detector design of the Investigating Solar Axion by Iron-57 (ISAI) is shown in Fig.~\ref{design}. 
ISAI consists of two detector modules.
In one module, a 95.85\% enriched $^{57}$Fe foil, whose size is 3.2\,cm~$\times$~3.2\,cm~$\times$~40\,$\mu$m, is set and used as the axion absorber.
In the other module, a natural Fe foil is set instead and used as a reference for the background study.
The escape probability of 14.4\,keV X-rays from such foil is 24.8\%. Foils are covered by two XRPIX detectors, whose detection area and the depletion thickness are 21.9\,mm $\times$ 13.8\,mm and 300\,$\mu$m, respectively.
The distance between the XRPIXs and the foil is 3\,mm and its detection efficiency of the 14.4\,keV X-rays is 14.5\%.
Oxygen-free copper plates with a thickness of  5\,mm are arranged around the XRPIXs as passive shieldings to protect from X- and beta-rays emitted from the lead shieldings outside. 
The lead shieldings, whose thickness is 50\,mm are specified to reduce the penetration of environment gamma rays.
Anti-counters made by plastic scintillators are used to measure the trajectory of cosmic-ray particles to reduce these backgrounds.
These instruments are located inside a thermostatic chamber to keep the XRPIXs at a low temperature.

\subsection{XRPIX}
\begin{figure}[!h]
\centering
\includegraphics[width=\textwidth]{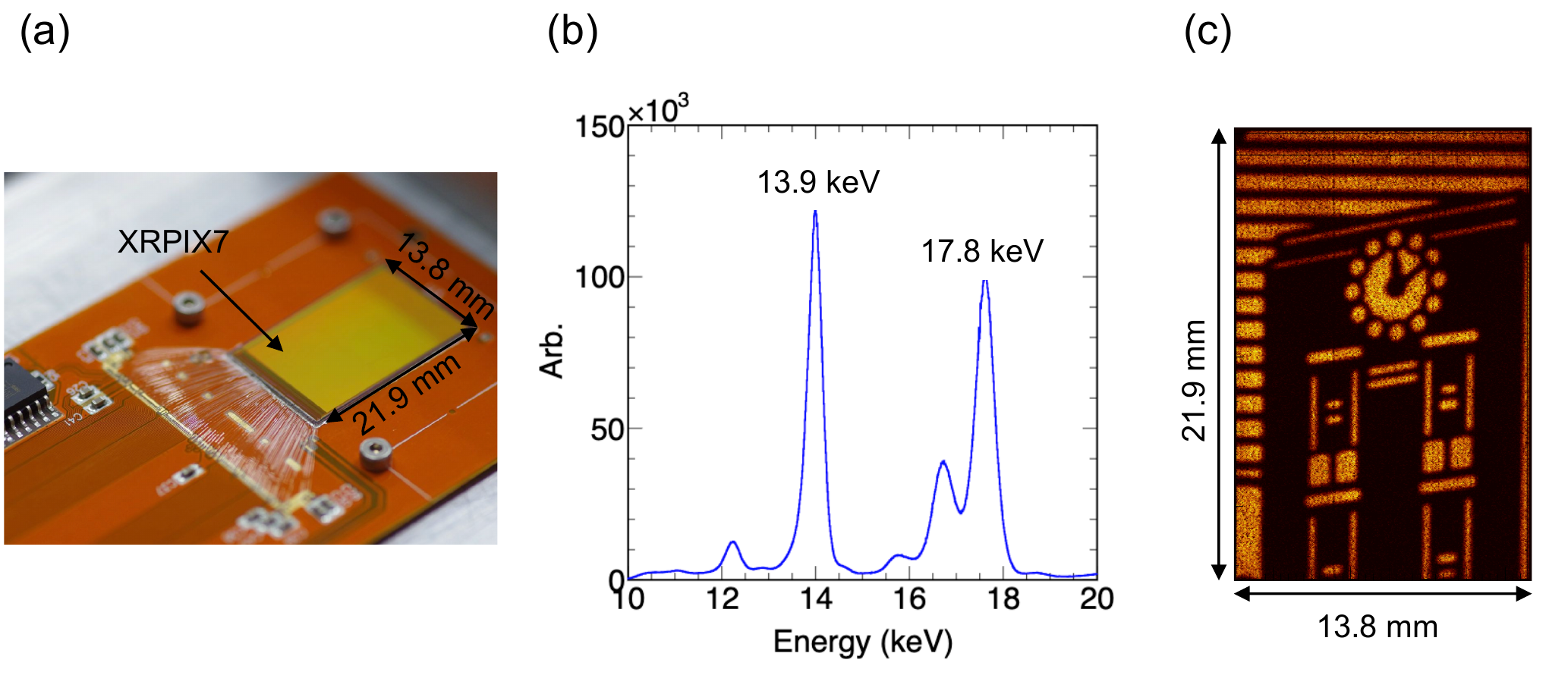}
\caption{(a)~Photograph of XRPIX7. (b)~Spectrum of $^{241}$Am with event-driven readout mode. (c)~X-ray image using the metal mask. }
\label{xrpix}
\end{figure}

XRPIX\cite{5975231,6412755,6407484,Takeda_2015,HARADA2019468,HAYASHI2019400,10.1117/12.2312098}, which is one of the monolithic active pixel detectors based on the silicon-on-insulator (SOI) pixel technology, has been developed for a future X-ray astronomy mission, so-called FORCE~\cite{10.1117/12.2309344}, since 2011. We are planning to employ series number seven (XRPIX7) which has the largest detection area.
The XRPIX7 has 608~$\times$~384 pixels and the pixel size is 36\,$\mu$m~$\times$~36\,$\mu$m. Therefore, the XRPIX7 covers a detection area of 21.9\,mm $\times$ 13.8\,mm.
The picture of XRPIX7 is shown in Fig.~\ref{xrpix}(a).
Each pixel is equipped with a CMOS circuit for the signal processing, including comparators. 
The comparator gives the hit trigger, timing, and position. 
Furthermore, the XRPIX7 outputs the pulse heights around the hit pixels as the event-driven readout mode. 
Therefore, the XRPIX7 offers a time resolution better than 10\,$\mu$s, $\sim 10^{3-5}$ faster than scientific CCDs, and enables the use of an anti-coincidence technique to reduce backgrounds. 
Fig.~\ref{xrpix}(b) shows the observed spectrum of an $^{241}$Am source with the event-driven readout mode.
The obtained energy resolution was 478\,eV (FWHM) at 13.9\,keV on the whole detection area. An X-ray image taken with a metal mask is shown in Fig.~\ref{xrpix}(c).

\ikeda{The conventional circuit board made of G10 glass epoxy included many background sources of $^{238}$U, $^{232}$Th, and $^{40}$K~\cite{ONUKI2019448}. Hence, we developed a rigid flexible print circuit board as a low-background readout system (Fig.~\ref{xrpix}(a)). The XRPIX7 bare chip is mounted on the flexible print circuit part without G10 glass epoxy and the rigid print circuit is  part from the sensor. Only ceramic capacitors are implemented near the XRPIX7. Compared to the conventional circuit board, the background rate in the energy region of 14.4\,keV is a factor of about 10$^{-3}$.}

\subsection{Plastic scintillator as anti-counter}
An anti-counter system comprised of plastic scintillators is installed on the top and bottom of the main detector as shown in Fig.~\ref{design}. 
The cross-section of each scintillator is not a square, but an isosceles triangle with a width of 5\,cm and a height of 1\,cm to enhance the position accuracy by a signal ratio of two adjacent scintillators. 
There are 11 scintillators with a length of 30\,cm in each layer. 
Two sets of two layers of X and Y directions are capable of measuring the trajectory of an individual cosmic ray particle as a position-sensitive anti-counter. 
The signals of scintillators are measured by SiPM (S13360-1375PE, Hamamatsu), and all of their waveforms are digitized by a dual 32ch 14-bit Flash ADC (ADS52J90, Texas Instrument).

\section{Sensitivity}
\begin{figure}[!h]
\centering
\includegraphics[width=8cm]{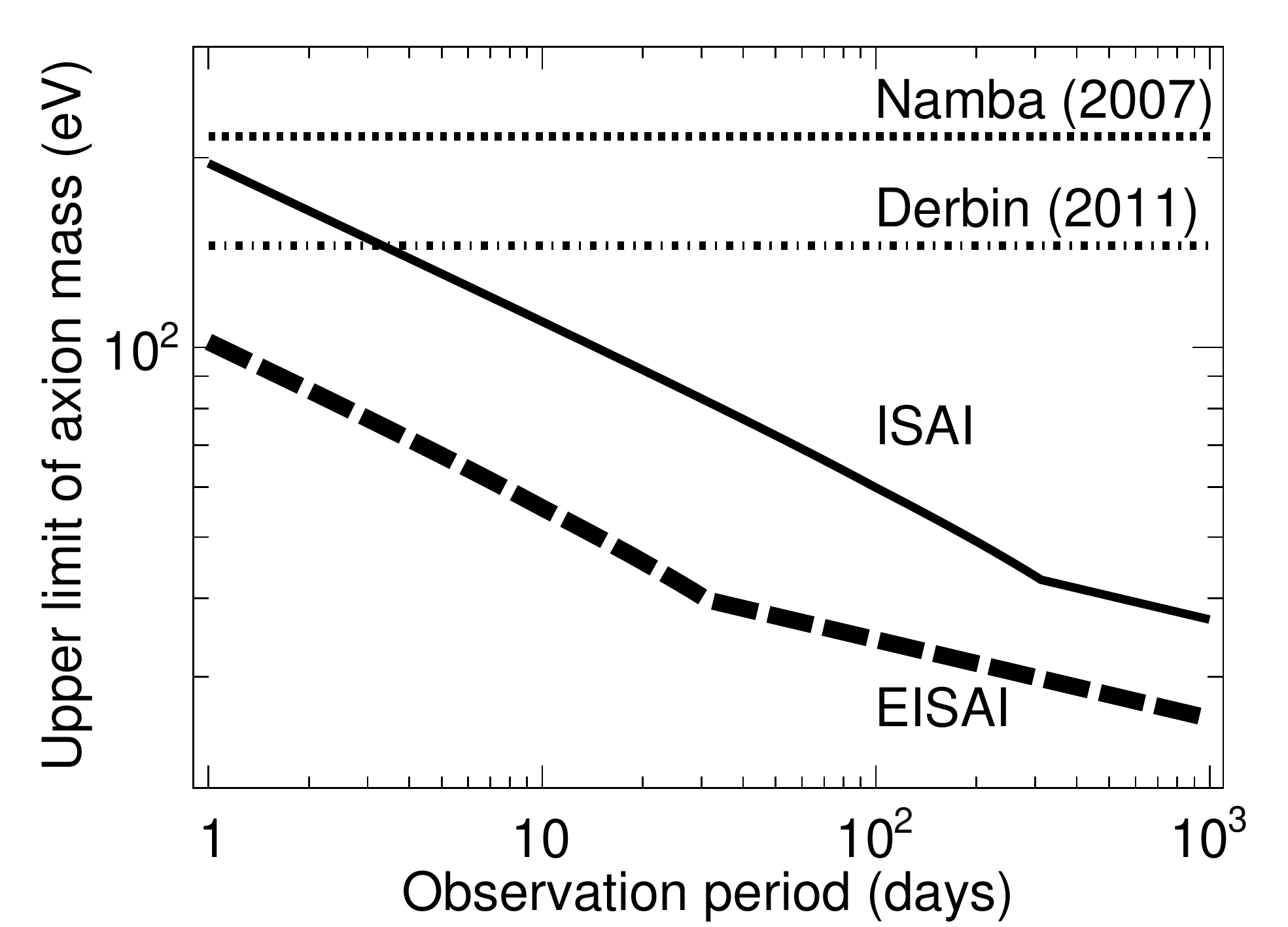}
\caption{Expected sensitivities. The solid and dashed lines show this work and future work of 10 times $^{57}$Fe, respectively. The dotted and dash-dotted lines indicate the constraints from Namba~\cite{NAMBA2007398} and Derbin~\cite{Derbin}, respectively.}
\label{sensitivity}
\end{figure}

Using Eqs.~(\ref{eq:mass}) and (\ref{eq:rate}), the axion mass is computed via the following equation:
\begin{equation}
    m_{a} = 5.55 \times \biggl( \frac{R}{1 ~{\rm day^{-1}kg^{-1}}} \biggr)^{1/4} ~{\rm eV},
\end{equation}
where $R$ is the $^{57}$Fe de-excitation rate and rewritten as $R=N_{\gamma}/(M\eta\epsilon)$. We assumed the mass ratio $z$ and the nuclear structure $C$ as 0.56 and 0.27, respectively. Here, $N_{\gamma}$ is the count rate, $M$ is the  $^{57}$Fe mass, $\eta$ is the emission probability of 14.4~keV X-rays, and $\epsilon$ is the detection efficiency. 
\ikeda{The sensitivity is calculated by replacing $N_{\gamma}$ with the background rate $N_{\rm BG}$. The black solid line in Fig.~\ref{sensitivity} shows the expected sensitivity where we assumed that $M=127$~mg, $\eta=0.105$, and $\epsilon=0.145$. Furthermore, $N_{\rm BG}=0.004$~counts/day was estimated by the internal background simulation associated with the energy resolution of 250~eV (FWHM) at 14.4~keV.}
A 95\% confidence level  upper limit on the axion mass of 145~eV corresponding to Derbin's result\cite{Derbin} is reachable in about four days. The dashed line indicates the expected sensitivity of the future work with ten times the target mass.

\section{Future prospects}
We plan to start the commissioning run without the anti-counter and the $^{57}$Fe foil at the end of 2022. Then, the primary performance of XRPIX7 and the ambient gamma-ray background are verified over three months. Anti-counters are calibrated using muons of cosmic rays in parallel. In 2023, a scientific run will be carried out with the complete system of ISAI for one year. While the current energy resolution of XRPIX7 reaches less than 250 eV, we will be able to record the best upper limit in about one week. Moreover, we are developing the new XRPIX of series number ten to overcome the energy resolution. The XRPIX7 will be replaced with XRPIX10 as soon as development is complete.



\section*{Acknowledgements}
This work was partly supported by Japan Society for the promotion of Science (JSPS), KAKENHI Grant-in-Aids for Scientific Research, Grant Number JP 21K18151, 22H04572, 21H05461, 21H04493 and JSPS Research Fellow, Grant Number JP 22J00064.
The authors thank S. Moriyama and T. Namba for productive and valuable discussions.

\bibliography{SciPost_Example_BiBTeX_File.bib}

\begin{thebibliography}{10}
\providecommand{\url}[1]{\texttt{#1}}
\providecommand{\urlprefix}{URL }
\expandafter\ifx\csname urlstyle\endcsname\relax
  \providecommand{\doi}[1]{doi:\discretionary{}{}{}#1}\else
  \providecommand{\doi}{doi:\discretionary{}{}{}\begingroup
  \urlstyle{rm}\Url}\fi
\providecommand{\eprint}[2][]{\url{#2}}

\bibitem{Crewther:1979pi}
R.~J. Crewther, P.~Di~Vecchia, G.~Veneziano and E.~Witten,
\newblock \emph{{Chiral Estimate of the Electric Dipole Moment of the Neutron
  in Quantum Chromodynamics}},
\newblock Phys. Lett. B \textbf{88}, 123 (1979),
\newblock \doi{10.1016/0370-2693(79)90128-X},
\newblock [Erratum: Phys.Lett.B 91, 487 (1980)].

\bibitem{Afach:2015sja}
J.~M. Pendlebury \emph{et~al.},
\newblock \emph{{Revised experimental upper limit on the electric dipole moment
  of the neutron}},
\newblock Phys. Rev. D \textbf{92}(9), 092003 (2015),
\newblock \doi{10.1103/PhysRevD.92.092003},
\newblock \eprint{1509.04411}.

\bibitem{PhysRevLett.38.1440}
R.~D. Peccei and H.~R. Quinn,
\newblock \emph{$\it{CP}$ conservation in the presence of pseudoparticles},
\newblock Phys. Rev. Lett. \textbf{38}, 1440 (1977),
\newblock \doi{10.1103/PhysRevLett.38.1440}.

\bibitem{PhysRevLett.40.223}
S.~Weinberg,
\newblock \emph{A new light boson?},
\newblock Phys. Rev. Lett. \textbf{40}, 223 (1978),
\newblock \doi{10.1103/PhysRevLett.40.223}.

\bibitem{Wilczek:1977pj}
F.~Wilczek,
\newblock \emph{{Problem of Strong $P$ and $T$ Invariance in the Presence of
  Instantons}},
\newblock Phys. Rev. Lett. \textbf{40}, 279 (1978),
\newblock \doi{10.1103/PhysRevLett.40.279}.

\bibitem{PhysRevLett.50.925}
J.~Ipser and P.~Sikivie,
\newblock \emph{Can galactic halos be made of axions?},
\newblock Phys. Rev. Lett. \textbf{50}, 925 (1983),
\newblock \doi{10.1103/PhysRevLett.50.925}.

\bibitem{Redondo_2013}
J.~Redondo,
\newblock \emph{Solar axion flux from the axion-electron coupling},
\newblock Journal of Cosmology and Astroparticle Physics \textbf{2013}(12), 008
  (2013),
\newblock \doi{10.1088/1475-7516/2013/12/008}.

\bibitem{CAST2017}
V.~Anastassopoulos, S.~Aune, K.~Barth, A.~Belov, H.~Br{\"a}uninger,
  G.~Cantatore, J.~M. Carmona, J.~F. Castel, S.~A. Cetin, F.~Christensen, J.~I.
  Collar, T.~Dafni \emph{et~al.},
\newblock \emph{New cast limit on the axion--photon interaction},
\newblock Nature Physics \textbf{13}(6), 584 (2017),
\newblock \doi{10.1038/nphys4109}.

\bibitem{PhysRevD.102.072004}
E.~Aprile, J.~Aalbers, F.~Agostini, M.~Alfonsi, L.~Althueser, F.~D. Amaro,
  V.~C. Antochi, E.~Angelino, J.~R. Angevaare, F.~Arneodo, D.~Barge, L.~Baudis
  \emph{et~al.},
\newblock \emph{Excess electronic recoil events in xenon1t},
\newblock Phys. Rev. D \textbf{102}, 072004 (2020),
\newblock \doi{10.1103/PhysRevD.102.072004}.

\bibitem{PhysRevLett.75.3222}
S.~Moriyama,
\newblock \emph{Proposal to search for a monochromatic component of solar
  axions using $^{57}\it{Fe}$},
\newblock Phys. Rev. Lett. \textbf{75}, 3222 (1995),
\newblock \doi{10.1103/PhysRevLett.75.3222}.

\bibitem{NAMBA2007398}
T.~Namba,
\newblock \emph{Results of a search for monochromatic solar axions using
  $^{57}\it{Fe}$},
\newblock Physics Letters B \textbf{645}(5), 398 (2007),
\newblock \doi{https://doi.org/10.1016/j.physletb.2007.01.005}.

\bibitem{5975231}
S.~G. Ryu, T.~G. Tsuru, S.~Nakashima, A.~Takeda, Y.~Arai, T.~Miyoshi,
  R.~Ichimiya, Y.~Ikemoto, H.~Matsumoto, T.~Imamura, T.~Ohmoto and A.~Iwata,
\newblock \emph{First performance evaluation of an x-ray soi pixel sensor for
  imaging spectroscopy and intra-pixel trigger},
\newblock IEEE Transactions on Nuclear Science \textbf{58}(5), 2528 (2011),
\newblock \doi{10.1109/TNS.2011.2160970}.

\bibitem{6412755}
S.~G. Ryu, T.~G. Tsuru, G.~Prigozhin, S.~Kissel, M.~Bautz, B.~LaMarr,
  S.~Nakashima, R.~F. Foster, A.~Takeda, Y.~Arai, T.~Imamura, T.~Ohmoto
  \emph{et~al.},
\newblock \emph{Tests with soft x-rays of an improved monolithic soi active
  pixel sensor},
\newblock IEEE Transactions on Nuclear Science \textbf{60}(1), 465 (2013),
\newblock \doi{10.1109/TNS.2012.2231880}.

\bibitem{6407484}
A.~Takeda, Y.~Arai, S.~G. Ryu, S.~Nakashima, T.~G. Tsuru, T.~Imamura, T.~Ohmoto
  and A.~Iwata,
\newblock \emph{Design and evaluation of an soi pixel sensor for trigger-driven
  x-ray readout},
\newblock IEEE Transactions on Nuclear Science \textbf{60}(2), 586 (2013),
\newblock \doi{10.1109/TNS.2012.2225072}.

\bibitem{Takeda_2015}
A.~Takeda, T.~Tsuru, T.~Tanaka, H.~Uchida, H.~Matsumura, Y.~Arai, K.~Mori,
  Y.~Nishioka, R.~Takenaka, T.~Kohmura, S.~Nakashima, S.~Kawahito
  \emph{et~al.},
\newblock \emph{Improvement of spectroscopic performance using a
  charge-sensitive amplifier circuit for an x-ray astronomical {SOI} pixel
  detector},
\newblock Journal of Instrumentation \textbf{10}(06), C06005 (2015),
\newblock \doi{10.1088/1748-0221/10/06/c06005}.

\bibitem{HARADA2019468}
S.~Harada, T.~G. Tsuru, T.~Tanaka, H.~Uchida, H.~Matsumura, K.~Tachibana,
  H.~Hayashi, A.~Takeda, K.~Mori, Y.~Nishioka, N.~Takebayashi, S.~Yokoyama
  \emph{et~al.},
\newblock \emph{Performance of the silicon-on-insulator pixel sensor for x-ray
  astronomy, xrpix6e, equipped with pinned depleted diode structure},
\newblock Nuclear Instruments and Methods in Physics Research Section A:
  Accelerators, Spectrometers, Detectors and Associated Equipment \textbf{924},
  468 (2019),
\newblock \doi{https://doi.org/10.1016/j.nima.2018.09.127},
\newblock 11th International Hiroshima Symposium on Development and Application
  of Semiconductor Tracking Detectors.

\bibitem{HAYASHI2019400}
H.~Hayashi, T.~G. Tsuru, T.~Tanaka, H.~Uchida, H.~Matsumura, K.~Tachibana,
  S.~Harada, A.~Takeda, K.~Mori, Y.~Nishioka, N.~Takebayashi, S.~Yokoyama
  \emph{et~al.},
\newblock \emph{Evaluation of kyoto’s event-driven x-ray astronomical soi
  pixel sensor with a large imaging area},
\newblock Nuclear Instruments and Methods in Physics Research Section A:
  Accelerators, Spectrometers, Detectors and Associated Equipment \textbf{924},
  400 (2019),
\newblock \doi{https://doi.org/10.1016/j.nima.2018.09.042},
\newblock 11th International Hiroshima Symposium on Development and Application
  of Semiconductor Tracking Detectors.

\bibitem{10.1117/12.2312098}
T.~G. Tsuru, H.~Hayashi, K.~Tachibana, S.~Harada, H.~Uchida, T.~Tanaka,
  Y.~Arai, I.~Kurachi, K.~Mori, A.~Takeda, Y.~Nishioka, N.~Takebayashi
  \emph{et~al.},
\newblock \emph{{Kyoto's event-driven x-ray astronomy SOI pixel sensor for the
  FORCE mission}},
\newblock In A.~D. Holland and J.~Beletic, eds., \emph{High Energy, Optical,
  and Infrared Detectors for Astronomy VIII}, vol. 10709, p. 107090H.
  International Society for Optics and Photonics, SPIE,
\newblock \doi{10.1117/12.2312098} (2018).

\bibitem{10.1117/12.2309344}
K.~Nakazawa, K.~Mori, T.~G. Tsuru, Y.~Ueda, H.~Awaki, Y.~Fukazawa, M.~Ishida,
  H.~Matsumoto, H.~Murakami, T.~Okajima, T.~Takahashi, H.~Tsunemi
  \emph{et~al.},
\newblock \emph{{The FORCE mission: science aim and instrument parameter for
  broadband x-ray imaging spectroscopy with good angular resolution}},
\newblock In J.-W.~A. den Herder, S.~Nikzad and K.~Nakazawa, eds., \emph{Space
  Telescopes and Instrumentation 2018: Ultraviolet to Gamma Ray}, vol. 10699,
  p. 106992D. International Society for Optics and Photonics, SPIE,
\newblock \doi{10.1117/12.2309344} (2018).

\bibitem{ONUKI2019448}
Y.~Onuki, J.~A.~M. Grimaldo, T.~Ose, H.~Aihara, Y.~Inoue, Y.~Kamiya,
  K.~Shimazoe, T.~G. Tsuru, T.~Tanaka, K.~Miuchi, A.~Takeda and Y.~Arai,
\newblock \emph{Studies of radioactive background in soi pixel detector for
  solar axion search experiment},
\newblock Nuclear Instruments and Methods in Physics Research Section A:
  Accelerators, Spectrometers, Detectors and Associated Equipment \textbf{924},
  448 (2019),
\newblock \doi{https://doi.org/10.1016/j.nima.2018.07.056},
\newblock 11th International Hiroshima Symposium on Development and Application
  of Semiconductor Tracking Detectors.

\bibitem{Derbin}
A.~V. Derbin, V.~N. Muratova, D.~A. Semenov and E.~V. Unzhakov,
\newblock \emph{New limit on the mass of 14.4-kev solar axions emitted in an m1
  transition in $^{57}\it{Fe}$ nuclei},
\newblock Physics of Atomic Nuclei \textbf{74}(4), 596 (2011),
\newblock \doi{10.1134/S1063778811040041}.

\end{thebibliography}

\nolinenumbers

\end{document}